\documentstyle[aps,epsfig,amstex,twocolumn]{revtex}
\newcommand{\ve}[1]{{\mathbf{#1}}}
\begin{document}
\title{Low energy Collective Modes of a Superfluid Trapped atomic Fermi Gas}

\author{G.\ M.\ Bruun and B.\ R.\ Mottelson}
\address{Nordita, Blegdamsvej 17, 2100 Copenhagen, Denmark}
\maketitle

\begin{abstract} 
We consider the low energy collective mode spectrum of a superfluid Fermi gas in a spherical trap in the collisionless regime.
Using a self-consistent random-phase approximation, the effects of superfluidity on modes of dipole and quadrupole symmetries 
 are systematically
 examined. The spectrum is calculated  for varying  pairing strength and temperature and we identify several spectral features
 such as the emergence of Goldstone modes that can be used to detect the onset of superfluidity. 
 Our analysis is relevant for present experiments aimed at   observing a
superfluid phase transition in trapped Fermi gases.
\end{abstract}

Pacs Numbers: 32.80.Pj, 05.30.Fk, 67.55.Jd 
\
\

Properties of trapped Fermi gases are attracting increasing attention in the field of ultracold atomic systems. 
Experimentally, the  trapping and cooling of fermionic alkalis has been demonstrated reaching temperatures 
as low as $\sim T_F/4$   for  $^{40}$K~\cite{DeMarco} and $^6$Li~\cite{Truscott,Schreck,O'Hara} with 
$T_F$ denoting the Fermi temperature.
One goal of the present experimental effort is to observe a transition to a superfluid phase at some critical 
temperature $T_c$ for systems with an attractive interaction between atoms in two different hyperfine states~\cite{StoofBCS}.
Several proposals for the  detection of this phase transition using  
the interaction with light~\cite{Lscattering},  thermodynamic and various collective and quasi particle properties~\cite{Zambelli00} 
have been presented.

The purpose of this  paper is to examine the effects of superfluidity on the low energy collective mode 
spectrum. The  modes can  be measured with high precision as shape 
oscillations of the atomic cloud and 
 such studies have proven to be very useful in the field of Bose-Einstein condensation~\cite{Edwards1}. 
Ultracold atomic Fermi gases  provide an experimental realization of a neutral superfluid 
with a simple tunable interaction. Both regimes where the pairing is a small perturbation on the normal properties of the gas 
and where it dominates the low frequency response could be experimentally obtainable.
 Also, finite temperature effects should  be measurable. 
This makes the study of collective modes in these systems interesting as they in some sense are simpler and easier to 
experimentally manipulate than other well known superfluid Fermi systems such as  liquid $^3$He and the atomic nucleus. 

We present an  analysis of the low energy collective mode spectrum of a trapped superfluid Fermi gas
 for various  temperatures and coupling strengths. The trapping potential has a qualitative effect
on the low energy modes as opposed to the high energy regime, where one can use the well-known results for a homogeneous 
superconductor~\cite{Minguzzi}. 
We assume that the underlying normal gas is in
 the collisionless limit such that   the lifetime $\tau$ of the quasi particles is much longer
than the characteristic period of motion ({\it{i.e.}} $\omega_T\tau\gg 1$
for atoms in a harmonic trap of frequency $\omega_T$). It is shown that  superfluidity has qualitative effects on the 
modes of dipole and quadrupole symmetries  and we thus propose several ways of detecting the superfluid
 transition. An essential difference between the density and spin-density fluctuation modes is demonstrated and 
we also analyze the effects of a non-zero temperature.

Consider a gas of fermionic atoms of mass $m$ in two hyperfine states 
 $|\sigma=\uparrow,\downarrow\rangle$ trapped by a spherically 
symmetric harmonic potential $U_0(\ve{r})=\frac{1}{2}m\omega_T^2r^2$. 
The numbers $N_\sigma$ of atoms trapped in each hyperfine state are assumed to be equal as this is the 
optimum situation for Cooper pairing. The quasi particle excitations $\eta$ of the system are calculated by 
 the Hartree-Fock-Bogoliubov (HFB) method.  Using  a zero-range  pseudo-potential method to describe the
interaction between the atoms,  the Bogoliubov-de Gennes (BdG) equations become~\cite{deGennes}:
\begin{eqnarray}\label{BdGeqn}
E_\eta u_\eta({\mathbf{r}}) &=& [{\cal H}_0 + W({\mathbf{r}})
 u_\eta({\mathbf{r}}) +\Delta({\mathbf{r}}) v_\eta({\mathbf{r}})\nonumber \\
E_\eta v_\eta({\mathbf{r}}) &=& -[{\cal H}_0 + W({\mathbf{r}})]v_\eta({\mathbf{r}}) +
\Delta({\mathbf{r}}) u_\eta({\mathbf{r}})\label{BdG}.
\end{eqnarray}
Here, ${\mathcal{H}}_0= -(\hbar^2/2m)\nabla^2+U_0({\mathbf{r}})-\mu_{\rm F}$,
 $W({\mathbf{r}})\equiv g\langle\hat{\psi}_{\sigma}^{\dagger}({\mathbf{r}})
\hat{\psi}_{\sigma}({\mathbf{r}})\rangle=g\langle\hat{\rho}_\sigma({\mathbf{r}})\rangle$ 
is the Hartree potential, and 
 $\hat{\psi}_{\sigma} ({\mathbf{r}})$ is the field operator that annihilates an atom in spin state 
 $\sigma$ at position   $\ve{r}$.  The coupling constant  is 
 $g=4\pi a\hbar^2/m$ with $a$ being the $s$-wave scattering length between atoms in the two 
different hyperfine states ($a<0$, i.e. attraction) 
and $\mu_F$ denotes the chemical potential. The pairing field is defined as
 $\Delta({\mathbf{R}})\equiv -g\lim_{r\rightarrow 0}\partial_r[r\langle\hat{\psi}_\uparrow({\mathbf{R}}+{\mathbf{r}}/2)
\hat{\psi}_\downarrow({\mathbf{R}}-{\mathbf{r}}/2)\rangle]$~\cite{BruunBCS}.
The quasi particles with energies $E_\eta$ are described
 by the Bogoliubov wave  functions $u_\eta({\mathbf{r}})$ and $v_\eta({\mathbf{r}})$. 

We are  in this paper interested in the response of the gas in the 
superfluid phase to various 
particle conserving time-dependent external perturbations of the kind 
 $\hat{F}(t)\propto\exp(i\omega t)\sum_\sigma\int d^3rF_\sigma(\ve{r})\hat{\rho}_\sigma(\ve{r})$.
In the collisionless limit, the appropriate 
framework to calculate the collective mode spectrum is the self-consistent random phase approximation generalized to
superfluid systems~\cite{Anderson,Martin}. The linear response of the superfluid is characterized by a 
matrix consisting of two-particle correlation functions:
\begin{equation}\label{supermatrix}
\Pi(\omega) =\left\{
\begin{array}{cccc}
\langle\langle\hat{\rho}_\uparrow\hat{\rho}_\uparrow\rangle\rangle &
\langle\langle\hat{\rho}_\uparrow\hat{\rho}_\downarrow\rangle\rangle&
\langle\langle\hat{\rho}_\uparrow\hat{\chi}\rangle\rangle&
\langle\langle\hat{\rho}_\uparrow\hat{\chi}^{\dagger}\rangle\rangle\\
\langle\langle\hat{\rho}_\downarrow\hat{\rho}_\uparrow\rangle\rangle &
\langle\langle\hat{\rho}_\downarrow\hat{\rho}_\downarrow\rangle\rangle&
\langle\langle\hat{\rho}_\downarrow\hat{\chi}\rangle\rangle&
\langle\langle\hat{\rho}_\downarrow\hat{\chi}^{\dagger}\rangle\rangle\\
\langle\langle\hat{\chi}\hat{\rho}_\uparrow\rangle\rangle &
\langle\langle\hat{\chi}\hat{\rho}_\downarrow\rangle\rangle&
\langle\langle\hat{\chi}\hat{\chi}\rangle\rangle&
\langle\langle\hat{\chi}\hat{\chi}^{\dagger}\rangle\rangle\\
\langle\langle\hat{\chi}^\dagger\hat{\rho}_\uparrow\rangle\rangle &
\langle\langle\hat{\chi}^\dagger\hat{\rho}_\downarrow\rangle\rangle&
\langle\langle\hat{\chi}^\dagger\hat{\chi}\rangle\rangle&
\langle\langle\hat{\chi}^\dagger\hat{\chi}^{\dagger}\rangle\rangle
\end{array}
\right\}.
\end{equation}
Here,  $\langle\langle \hat{A}\hat{B}\rangle\rangle$ is the Fourier transform of the 
retarded function  $-i\Theta(t-t')\langle[\hat{A}(\ve{r},t),\hat{B}(\ve{r}',t')]\rangle$ and $\langle \ldots\rangle$
denotes the thermal average. The operator 
 $\hat{\chi}=\hat{\psi}_\downarrow(\ve{r})\hat{\psi}_\uparrow(\ve{r})$
 describes the local fluctuations in the pairing field. Note that we have split the correlation functions 
into the different hyperfine components since we also wish to consider spin modes 
in which the
 two hyperfine species oscillate in anti-phase. For the normal phase, the 
density and spin-density fluctuation modes can be obtained 
 by calculating the upper left quadrant of $\Pi(\omega)$ in Eq.(\ref{supermatrix})~\cite{BruunCollN}. 
For the superfluid phase however,   fluctuations in the pairing field are coupled to  density fluctuation modes
through the   non-particle-conserving correlation functions such as 
 $\langle\langle\rho\chi\rangle\rangle$ and we need to consider the full matrix  in Eq.(\ref{supermatrix}).
Within RPA, we have
\begin{equation}\label{Inversion}
\Pi(\omega) =[1-\Pi_0(\omega){\mathcal{G}}]^{-1}\Pi_0(\omega)
\end{equation}
with 
\begin{equation}
{\mathcal{G}}=\frac{g}{\hbar}\delta(\ve{r}-\ve{r}')\left\{
\begin{array}{cccc} 0&1&0&0\\1&0&0&0\\0&0&0&1\\0&0&1&0
\end{array}
\right\}
\end{equation}
describing the interaction between quasi particles in opposite hyperfine states.
Here the matrix products denote: $AB\equiv\int d^3r''A(\ve{r},\ve{r}'')B(\ve{r}'',\ve{r}')$
and $\Pi_0(\omega)$ is the matrix in Eq.(\ref{supermatrix}) calculated for non-interacting excitations 
using the HFB approximation given by Eq.(\ref{BdGeqn}).

For a spherically symmetric trap, the correlation functions split into terms describing the 
various multipole modes~\cite{Bertsch}. For instance, we obtain for the correlation function 
 $\langle\langle\hat{\rho}_\sigma(\ve{r})\hat{\rho}_\sigma(\ve{r}')\rangle\rangle$:
\begin{equation}\label{Angular}
\langle\langle\hat{\rho}_\sigma(\ve{r})\hat{\rho}_\sigma(\ve{r}')\rangle\rangle=
\sum_{LM}\Xi_L(r,r',\omega)Y_{LM}(\theta,\phi)Y^*_{LM}(\theta',\phi')
\end{equation}
with $r$ denoting the  distance from the center of the trap and the $Y_{LM}(\theta,\phi)$ 
 the usual spherical harmonics. The expressions for the other correlation functions appearing in Eq.(\ref{supermatrix})
are completely equivalent. Using Wicks theorem, we express the  independent particle correlation functions 
$\langle\langle\ldots\rangle\rangle_0$
used to form $\Pi_0(\omega)$ in terms the quasi particle  energies/wave functions obtained from
 Eq.(\ref{BdGeqn}). For instance, for 
$\langle\langle\hat{\rho}_\sigma(\ve{r})\hat{\rho}_\sigma(\ve{r}')\rangle\rangle_0$, we obtain 
\begin{gather}\label{densdens}
{\Xi_0}_L(r,r',\omega)=
\sum_{ll'}\frac{\hbar(2l+1)(2l'+1)}{4\pi(2L+1)r^2{r'}^2}|\langle ll'00|L0\rangle|^2\nonumber\\
\sum_{\eta\eta'}\left[\frac{uuuu(f-f')}{E-E'-\hbar\omega-i\delta}+\frac{uuvv(f+f'-1)}{E+E'-\hbar\omega-i\delta}\right.
\nonumber\\
\left.-\frac{vvuu(1-f-f')}{E+E'+\hbar\omega+i\delta}+\frac{vvvv(f'-f)}{-E+E'-\hbar\omega-i\delta}\right].
\end{gather}
Here, we write the Bogoliubov wave functions from Eq.(\ref{BdGeqn}) with angular momentum quantum 
numbers $lm$ and energy $E_{\eta l}$ in the form 
$u_{\eta lm}({\mathbf{r}})= r^{-1}u_{\eta l}(r)Y_{lm}(\theta,\phi)$ and 
$v_{\eta lm}({\mathbf{r}}) = r^{-1}v_{\eta l}(r)Y_{lm}(\theta,\phi)$.  We define
$uuvv\equiv  u_{\eta l}(\ve{r})u_{\eta l}(\ve{r}')v_{\eta' l'}(\ve{r})v_{\eta' l'}(\ve{r}')$ etc., and  
$f=[\exp(\beta E_{\eta l})+1]^{-1}$ is the Fermi function with $\beta=1/k_BT$. The Clebsch-Gordan coefficients
 $\langle ll'00|L0\rangle$ yield strong selection rules for $l$ and $l'$: for the monopole modes ($L=0$) one has
 $l=l'$, for the dipole mode ($L=1$) $l'=l\pm1$ etc.

There is one last technical issue to resolve: The correlation 
function $\langle\langle\chi(\ve{r})\chi^\dagger(\ve{r}')\rangle\rangle$ is ultraviolet divergent
 when expressed as a sum over 
Bogoliubov wave functions as in Eq.(\ref{densdens})~\cite{Leggett}. Using the pseudo-potential method in a way similar to
 that  described in Ref.\ \cite{BruunBCS}, this divergence is removed by the substitution
\begin{equation}
\langle\langle\chi(\ve{r})\chi^\dagger(\ve{r}')\rangle\rangle\rightarrow
\langle\langle\chi(\ve{r})\chi^\dagger(\ve{r}')\rangle\rangle+\frac{1}{2}\delta(\ve{r}-\ve{r}')G^{irr}_{\mu_F}(\ve{r}).
\end{equation}
Here $G^{irr}_{\mu_F}(\ve{r})$ is the part of the  the single particle Greens function 
 $G_{\mu_F}(\ve{r},\ve{x})\equiv\langle\ve{r}+\ve{x}/2|{\mathcal{H}}_0^{-1}|\ve{r}+\ve{x}/2\rangle$
that diverges as $1/x$ for $x\rightarrow0$~\cite{BruunBCS}. One can show that for the homogeneous case, this 
procedure is equivalent to subtracting the non-regularized gap equation from the calculation 
of $\langle\langle\chi\chi^\dagger\rangle\rangle$ and the diagonalization of $\Pi(\omega)$ then yields the 
well-known Bogoliubov-Anderson mode~\cite{Anderson}.

The structure of the self-consistent RPA calculation is then the usual one: First we obtain a self-consistent solution 
to the BdG equations given by Eq.(\ref{BdGeqn});
 then  $\Pi_0(\omega)$ is formed using the obtained Bogoliubov wave functions/energies;
 and finally the RPA response function is calculated from Eq.\ (\ref{Inversion}). 
 The poles of $\Pi(\omega)$ then give the collective modes of the gas. 

The quantity of experimental interest is the strength function 
\begin{equation}
S(F,\omega)= \sum_{nm}\frac{e^{-\beta E_n}-e^{-\beta E_m}}{\mathcal{Z}}|\langle n|\hat{F}|m\rangle|^2
\delta(\hbar\omega+E_n-E_m)
\end{equation}
which is directly related to the net transitions per unit time with energy $\hbar\omega$ induced by the operator
 $\hat{F}$. 
Here ${\mathcal{Z}}$ is the grand partition function and $|n\rangle$ is an eigenstate of the Hamiltonian with 
energy $E_n$. For operators of the form 
$\hat{F}(t)\propto\exp(i\omega t)\sum_\sigma\int d^3rF_\sigma(\ve{r})\hat{\rho}_\sigma(\ve{r})$, it is given by  
\begin{equation}\label{strength}
S(F,\omega)=-\frac{1}{\hbar\pi}\sum_{\sigma,\sigma'}\int d^3rd^3r'F_\sigma F'_{\sigma'}
Im[\langle\langle\hat{\rho}_\sigma\hat{\rho}'_{\sigma'}\rangle\rangle].
\end{equation}
We shall calculate it for $F_\sigma(\ve{r})=F_\sigma(r)Y_{LM}(\theta,\pi)$
with  $F_\uparrow(r)=F_\downarrow(r)=r^2$ for the quadrupole ($L=2$) density 
mode.  For the  dipole symmetry ($L=1$), we  take $F_\uparrow(r)=-F_\downarrow(r)=r$ exciting the lowest spin-dipole mode. 
Taking $F_\uparrow(r)=F_\downarrow(r)=r$ for $L=1$ would simply excite the center-of-mass mode at $\omega=\omega_T$.
Taking $L=0$ excites modes of monopole symmetry. The influence of superfluidity on the monopole modes is somewhat 
complicated and will be analyzed in a future publication.

%An appealing characteristic of the RPA, is that it satisfied satisfies the $f$-sum rule. It reads~\cite{Bertsch}
%\begin{equation}\label{sumrule}
%\int d\omega S(F,\omega)\omega=\frac{1}{m}\sum_\sigma\int d^3r|\nabla F_\sigma(\ve{r})|^2
%\rho_\sigma(\ve{r}),
%\end{equation}
% $\rho_\sigma(\ve{r})=\langle\hat{\rho}_\sigma(\ve{r})\rangle$ being the equilibrium density. 
% Note that this sum rule is not obeyed if one uses the simple single particle response functions 
%determined by $\Pi_0(\omega)$ in Eq.(\ref{strength}) since simple BCS theory does not locally conserve the 
%number of particles. 
%Since the RPA calculation outlined above is rather involved with technical subtleties 
%such as the removal of ultraviolet divergencies, Eq.(\ref{sumrule}) provides a valuable check on the reliability
% of our results as $\rho_\sigma(r)$ is easily calculated from the solution to Eq.\ (\ref{BdGeqn}).

The properties of the superfluid gas in general depends  on quantities such as the size of the 
coherence length $\xi$ of the Cooper pairs, the quasi particle spectrum obtained by solving Eq.\ (\ref{BdGeqn}),  
 $\hbar\omega_T$ and the mean field splitting of the trap levels. For a large number of particles
 trapped and a strong interaction one has  $\Delta(r)\gg\hbar\omega_T$ for small $r$ and therefore $\xi(r)\ll R$
 with $R$ denoting the spatial extent of the cloud. 
 As we shall demonstrate, the pairing has strong effects on the low energy spectrum in this regime
and should thus be relatively easy to detect. We therefore believe this limit to be  of most relevance for the present
 experimental effort to observe the superfluid transition.   To illustrate some important physical concepts, we also present results 
for weaker coupling where  the pairing mainly occurs in the levels of the oscillator shell 
that is closest to the chemical potential (valence shell).  In this case, there are quasi particle excitations with 
$E_{\eta=0l}<\hbar\omega_T$ since the pairing energy is smaller than $\hbar\omega_T$.
 However, we do not in this paper consider the  theoretically interesting $\Delta\ll\hbar\omega_T$ limit 
in detail with $\Delta$ denoting the pairing energy. In this  regime the interplay between the shell structure of the trap and
 the Cooper pairing leads to a more complex phase diagram.

\begin{figure}
\centering
\epsfig{file=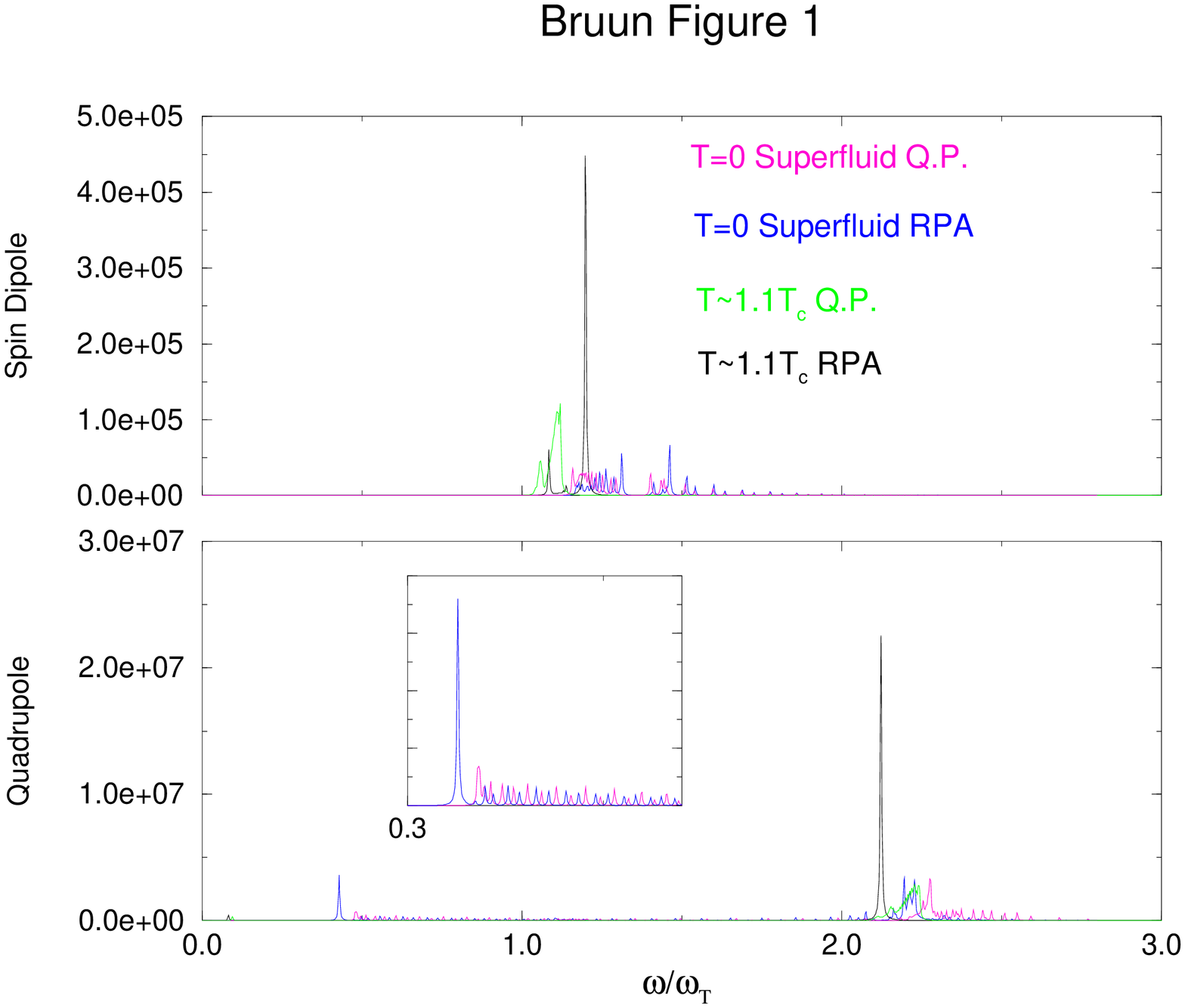,height=0.5\textwidth,width=0.28\textheight,angle=-90}
\caption{The collective (RPA) and quasi particle (Q.P.) response for the spin-dipole and quadrupole
 modes for $\Delta<\hbar\omega_T$}
\label{LowDelta}
\end{figure}

We first present typical results for a system with relatively weak pairing. 
 Figure (\ref{LowDelta}) exhibits  $S(F,\omega)$ (in trap units) given by Eq.\ (\ref{strength}) for the  spin-dipole, 
and quadrupole modes. There are $\sim 1.4\times 10^4$
 atoms trapped with a coupling strength such that the pairing field in the center of the trap 
is $\Delta(r=0)\simeq0.69\hbar\omega_T$ for $T=0$.  For the specific parameters chosen, it is mainly the levels 
within the harmonic oscillator shell  with principal quantum number $n=33$
 [with energy $\xi_n=(n+3/2)\hbar\omega_T$ in the 
non-interacting case] which Cooper  pair. The Hartree field has lowered the energy of this shell so that it is situated around 
the chemical potential at $\mu_F=32\hbar\omega_T$.
 We have added a small imaginary part $\Gamma=0.005\hbar\omega_T$ to the frequency
 to model a smooth response 
typically observed experimentally. For 
comparison we also plot  the response for $k_BT=\hbar\omega_T$ where the gas is in 
the normal phase ($k_BT_c\simeq0.9\hbar\omega_T$) and both the quasi particle and the collective (RPA)
response is shown. As expected, there is a large 
response at $\omega\simeq\omega_T, 2\omega_T$ for the  spin-dipole and quadrupole 
response respectively for the gas  in the normal phase. The shift of these  peaks away
 from the ideal gas result is, of course, a consequence of the particle-particle interaction~\cite{BruunCollN};  we  will 
not discus these shifts  further  in this paper as we are concentrating on features specific to superfluidity.

From Fig.\ (\ref{LowDelta}), we see that the presence of superfluidity introduces a qualitative new 
feature in the quadrupole response: a low frequency response for $\omega$ of the order $\sim 0.4\omega_T$.
The low energy response essentially comes from the  breaking of Cooper pairs in the lowest quasi particle shell.
 As pointed out by Baranov~\cite{Singleparticle}, 
the low energy quasi particle excitations [$E_{\eta=0,l}\ll\Delta(r=0)$]
are surface modes since the pair correlations imply that 
the interior of the paired system is an incompressible medium which 
expels the low energy quasi particles. Reflecting this effect, the
response observed at $\sim0.4\hbar\omega_T$ corresponds to a pair of 
quasi particles moving in the outer part of the cloud; the strength 
of this response in RPA is increased and the frequency decreased 
as a result of the (weak) coupling with the collective mode at 
$\sim2\hbar\omega_T$. 
This is illustrated by the inset  in Fig.\ (\ref{LowDelta})  which compares in detail the low energy single 
particle response calculated from $\Pi_0(\omega)$ with the collective response. The single 
particle response is generated by breaking of Cooper pairs due to excitations 
of the kind ${\hat{\gamma}}^\dagger_{\eta=0 l\uparrow}{\hat{\gamma}}^\dagger_{\eta'=0 l'\downarrow}|\Phi_0\rangle$
with ($l'=l,l\pm 2$) costing  $E_{\eta=0 l}+E_{\eta=0 l'}$ in energy. 
 Here, $|\Phi_0\rangle$ is the superfluid ground state and ${\hat{\gamma}}^\dagger_{\eta l\sigma}$
 creates a quasi particle with quantum numbers $\eta l\sigma$. The excitation energy associated with the
 lowest quasi particle states can mainly be attributed to the pair breaking  energy $\Delta$ and we write
  $E_{\eta=0l}+E_{\eta=0 l'}\sim2\Delta$. Thus, the low energy response consists of pair breaking excitations 
with  $\omega\sim2\Delta$ which is $\sim 0.4\hbar\omega_T$ for this set of parameters. As we shall see, 
this response is the precursor of a Goldstone mode present at stronger pairing. Furthermore, 
the peak at $\sim2\omega_T$ is fragmented and partially damped 
due to the pairing. We will demonstrate below
 that this mode simply disappears with the emergence of the Goldstone modes for 
stronger pairing.

For the spin-dipole mode, we see that the main effect of superfluidity in the $\Delta< \hbar\omega_T$ regime
is that it fragments and partially damps the normal phase response at $\sim\omega_T$. As for the quadrupole mode
at $\sim2\omega_T$, this effect is the precurser of the situation in the strong pairing regime, where we will show that 
there is no well-defined low frequency spin-dipole mode.

 Detailed effects of superfluidity on the mode spectrum in the weak pairing regime, such as the way the normal phase 
quadrupole/spin-dipole modes at $\sim2\omega_T/\omega_T$ are fragmented and 
damped with increasing pairing,  depend  sensitively on the specific 
parameters of the system such as the exact location of the chemical potential relative to the quasi particle bands.
 We will discuss these effects in detail
in a future publication.

\begin{figure}
\centering
\epsfig{file=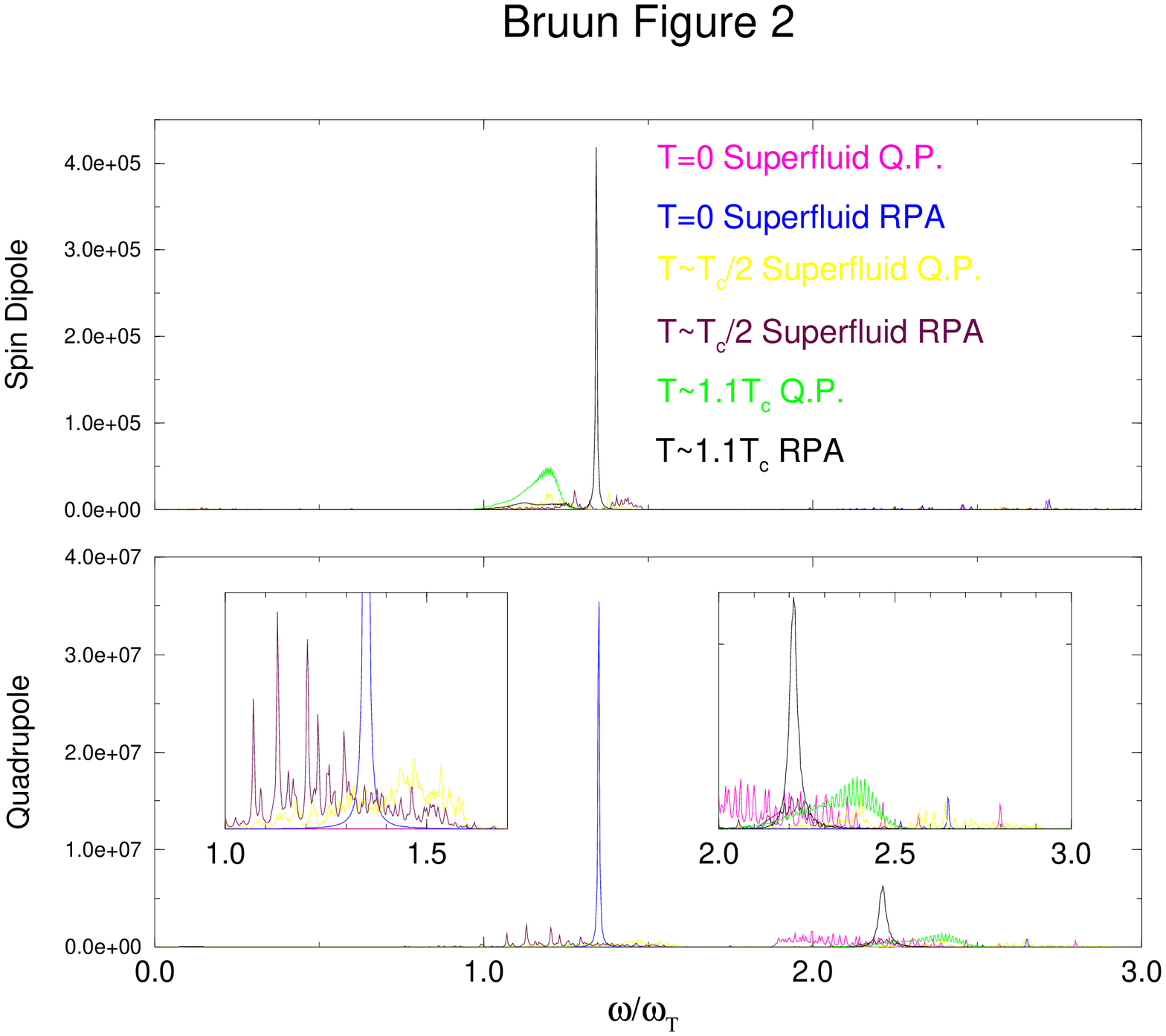,height=0.5\textwidth,width=0.28\textheight,angle=-90}
\caption{The collective (RPA) and quasi particle (Q.P.) response for the  spin-dipole and quadrupole
 modes for $\Delta>\hbar\omega_T$. The insets show various frequency regions in detail.}
\label{HighDelta}
\end{figure}

We now consider the stronger coupling limit when $\xi\ll R$.
 Figure (\ref{HighDelta}) exhibits  $S(F,\omega)$ for  $\sim 1.6\times 10^4$ atoms trapped with a coupling strength 
such that the pairing field in the center of the trap is $\Delta(r=0)=6\hbar\omega_T$ for $T=0$.   We have taken 
 $\Gamma=0.005\hbar\omega_T$ and the response is calculated for $T=0$ and $k_BT=1.4\hbar\omega_T$ 
 where the gas is still superfluid and $k_BT=3.0\hbar\omega_T$ ( where the gas is in the normal phase 
($k_BT_c\simeq2.8\hbar\omega_T$). Again, we plot both the quasi particle and the collective
response. 
A semi classical analysis shows that the lowest energy quasi particles are localized near the minimum 
of the function $\Delta(r)^2+[U_0(r)+W(r)+\hbar^2(l+1/2)^2/(2mr^2)-\mu_F]^2$~\cite{Singleparticle}. With increasing 
pairing, the width of this minimum narrows and the lowest quasi particle energy can be shown to increase as 
$\ln(\mu_F/\hbar\omega_T)$. For the specific parameters chosen above, the lowest quasi particle states have energies 
$E_{\eta=0l}\gtrsim\hbar\omega_T$.

From Fig.\ (\ref{HighDelta}), we see that for the quadrupole mode there is for $T=0$ now only one low frequency mode 
situated at  $\omega\simeq 1.4\omega_T$. This lower mode comes from the pairing degree of freedom 
[the lower right $2\times2$ part of the matrix $\Pi$ in Eq.(\ref{supermatrix})]; it is an example of the well-known Goldstone 
modes~\cite{Martin} with a finite frequency due to  the trapping potential. 
The low energy quasi particle response observed for weaker pairing [Fig.\ (\ref{LowDelta})]
has now become a collective (Goldstone) mode. For these modes, one can develop
a hydrodynamic theory valid for $\xi\ll R$~\cite{Martin} predicting a 
quadrupole  Goldstone mode at $\omega=\sqrt{2}\omega_T$~\cite{Baranov} in agreement with our 
result. Of experimental relevance is the temperature dependence of the spectrum. From Fig.\ (\ref{HighDelta}), we
see that as $T$  is lowered through $T_c$ the mode at $\omega\simeq2\omega_T$ is damped and a 
low frequency response emerges. At intermediate temperatures $0<T<T_C$,  the collective mode may suffer appreciably 
Landau damping [see the spectrum for $T=T_c/2$ in Fig.\ (\ref{HighDelta})]. The spectrum in Fig.\ (\ref{HighDelta})
is consistent with an interpretation in which the collective mode emerges as an undamped excitation 
as soon as the temperature is low enough so that the lowest quasi particle excitation energy is $\ge\sqrt{2}\hbar\omega_T$.
For the parameters in Fig.\ (\ref{HighDelta}), this occurs for $T\sim T_c/4$.

We  finally consider the low $\omega$ spin-dipole response. From Fig.\ (\ref{HighDelta}), we see that the normal 
phase mode at $\sim\omega_T$ is totally 
damped in the superfluid phase. This is to be expected for $\Delta>\hbar\omega_T$: The only kind of 
motion possible for the superfluid is potential flow, i.e.\ density fluctuation modes~\cite{Leggett}. All other
 modes such as spin-density modes requires the presence of a normal component which decreases with $T\rightarrow 0$. 
 The superfluid cannot participate in a mode of relative motion of the two hyperfine states; one needs to break pairs  and the
 response lies above the pairing energy. 
The damping of the spin-dipole mode can also be understood by writing the spin-dipole operator in terms of the 
quasi particle operators: 
 $[\hat{\rho}_\uparrow(\ve{r})-\hat{\rho}_\downarrow(\ve{r})]|\Phi_0\rangle\propto\sum_{\eta\eta'}
 [u_\eta(\ve{r})v_{\eta'}(\ve{r})-v_\eta(\ve{r})u_{\eta'}(\ve{r})]
\hat{\gamma}^\dagger_{\eta\uparrow} \hat{\gamma}^\dagger_{\eta'\downarrow}|\Phi_0\rangle$.
Pairing tends to mix a particle state with its time reversed hole-state yielding $u_\eta(\ve{r})\sim v_{\eta}(\ve{r})$ for a low energy 
excitation and therefore  $[u_\eta(\ve{r})v_{\eta'}(\ve{r})-v_\eta(\ve{r})u_{\eta'}(\ve{r})]\sim 0$. Thus, 
the low $\omega$ spin-dipole 
mode excited by an odd operator under time-reversal will  gradually disappear for  $T\rightarrow 0$ as can be seen from 
Fig.\ (\ref{HighDelta}). 

In conclusion, we have identified and discussed the physical origin of several clear-cut spectral features of superfluidity on
 the low frequency  collective mode
 spectrum of dipole and quadrupole symmetry in the collisionless regime for weak and strong pairing.
For a weak pairing $\Delta<\hbar\omega_T$, the main effect of superfluidity is the presence 
of a  low $\omega$ response for the quadrupole symmetry. Also, the 
 normal phase quadrupole and spin-dipole modes at $\sim2\omega_T$ and  $\sim\omega_T$ respectively are partially
damped.  For stronger pairing with  $\xi\ll R$, the Goldstone modes dominate the low
frequency spectrum and  the quadrupole mode is shifted to $\approx\sqrt{2}\omega_T$ in 
agreement with hydrodynamic theory. In this regime there is no well-defined low energy spin-dipole mode.
 As illustrated above, the emergence of these effects 
 should be straightforward to detect as one lowers $T$ through $T_c$ and we believe the results presented to be
relevant   for the present experimental effort related to  observing the predicted BCS transition.

We would like to acknowledge valuable discussions with C.\ J.\ Pethick.

\end{document}